\documentclass[11pt]{article}

\usepackage{fullpage}
\usepackage{epic,eepic,epsfig,graphics}
\usepackage{amsmath}
\usepackage{color}
\usepackage{latexsym}
\usepackage{tabularx}
\usepackage{times}
\usepackage{graphicx}
\usepackage{epstopdf}

{

\newtheorem{theorem}{Theorem}

\newcommand{\toto}{xxx}
\newenvironment{proofT}{\noindent{\bf
Proof }} {\hspace*{\fill}$\Box_{Theorem~\ref{\toto}}$\par\vspace{3mm}}
\newenvironment{proofL}{\noindent{\bf
Proof }} {\hspace*{\fill}$\Box_{Lemma~\ref{\toto}}$\par\vspace{3mm}}

\newenvironment{theorem-repeat}[1]{\begin{trivlist}
\item[\hspace{\labelsep}{\bf\noindent Theorem~\ref{#1} }]}%
{\end{trivlist}}

\newenvironment{lemma-repeat}[1]{\begin{trivlist}
\item[\hspace{\labelsep}{\bf\noindent Lemma~\ref{#1} }]}%
{\end{trivlist}}

\newcounter{linecounter}

\newcommand{\resetline}{\setcounter{linecounter}{0}}

\title{\bf Self-stabilizing Algorithm for Minimal $\alpha$-Dominating Set\footnote{The authors declare that they have no known competing for financial interests or personal relationships that could have appeared to influence the work reported in this paper.}} 

\author{Badreddine Benreguia and Hamouma Moumen\footnote{Correspondence: Dr. Hamouma Moumen. Email: hamouma.moumen@univ-batna2.dz} \\
Computer Science Dept., University of Batna 2\\
53, Road of Constantine, Fesdis\\
Batna 05078, Algeria\\
\{badreddine.benreguia, hamouma.moumen\}@univ-batna2.dz}

\date{}

\begin{document}

\maketitle

\begin{abstract}

A self-stabilizing algorithm for the minimal $\alpha$-dominating set is proposed in this paper. The $\alpha$-domination parameter has not used before in self-stabilization paradigm. Using an arbitrary graph with \textit{n} nodes and \textit{m} edges, the proposed algorithm converge in $O(nm)$ moves under distributed daemon. Simulation tests and mathematical proofs show the efficiency of the algorithm.\\

\noindent {\bf Keywords}: self-stabilizing algorithm; dominating set; $\alpha$-domination; distributed system; expression distance-2 model.

\end{abstract}


\section{Introduction}
Self-stabilization is a fault tolerance approach for distributed systems that has been introduced for the first time by Dijkstra \cite{Dij74}. A self-stabilizing distributed system is able to achieve a global correct configuration (without any external intervention), in a finite time, starting from an initial illegitimate configuration.
Various self-stabilizing distributed algorithms have been proposed in the literature using graph theory such as leader election, nodes coloring, domination problem, independent set identification, spanning tree construction. The reader can refer to the survey \cite{GK10} for more details of self-stabilizing algorithms.

Domination has been extensively studied in literature \cite{HHS98} and adopted in many real-life applications. It has been utilized for address routing, power management and clustering issues in ad-hoc networks \cite{LLY09,AWF03,BDTC04}. A dominating set is a subset $S$ of the graph nodes where every node is either in $S$ or is a neighbor of at least one node of $S$. The dominating set $S$ is minimal if there is no proper subset in $S$ that could be a dominating set. Recently, domination is used to influence (and change) the opinion of the users in the social networks  \cite{FEK09, AL18}. Before 2003, only greedy algorithms have been proposed in the literature to find (minimal) dominating sets. The first \textit{self-stabilizing} algorithm for (minimal) dominating set was proposed by Hedetniemi \textit{et al.} \cite{HHGS03}. After that, many variants of self-stabilising algorithms have been proposed imposing additional parameters of domination like total domination \cite{BYK14,GHJS03}, efficient domination \cite{Tur13,HHJKM12}, connected dominating set \cite{BBP13,DWS16}, influence domination \cite{WWTZ13,DWS14}, distance-k domination \cite{DDL18}. Each parameter has its benefits according to the used application. For example, connected dominating sets are generally used as backbone (infrastructure) in ad-hoc and sensor networks.

As for $\alpha$-domination concept (without self-stabilization concept), it has been studied for the first time by \cite{DHLM00}. Other results are given in \cite{DRV04} on $\alpha$-domination. Suppose that $G = (V, E)$ is a connected graph where $V$ is the set of nodes and $E$ is the set of edges. We say that $S  \subseteq V$ is $\alpha$-dominating if for all $v \in V-S, \frac{|N(v) \cap S|}{|N(v)|} \geq \alpha$, where $0 <\alpha \leq 1$ and $N(v)$ is the set of $v$ neighbors $i.e.$ $N(v)=\{u| vu \in E \}$. To the best of our knowledge, there is no self-stabilizing algorithm to find the $\alpha$-dominating set. Only, particular cases of $\alpha$-domination have been presented in the literature which are discussed in the next section.
  

\subsection{Related works of $\alpha$-domination on self-stabilization}
\label{Related works}

In self-stabilizing paradigm, few algorithms are proposed only for the particular instance of $\alpha = \frac{1}{2}$ (three works to the best of our knowledge). In these cases, authors try to find the minimal dominating set where each node (in $V-S$ or in $V$) is dominated by at least half ($\alpha = \frac{1}{2}$) of its neighborhood. Wang $et$ $al.$ have introduced the \textit{positive influence dominating set} \cite{FEK09}. A self-stabilizing algorithm known as MPIDS is presented for this parameter in \cite{WWTZ13}. We call $S \subseteq V$ a \textit{positive influence dominating set} if each node $v \in V$ is dominated by at least $\lceil \frac{|N(v)|}{2} \rceil$ (that is, $v$ has at least $\lceil \frac{|N(v)|}{2} \rceil$  neighbors in $S$). This algorithm can be considered as total $\frac{1}{2}$-domination because the condition of $\frac{1}{2}$-domination must be respected by all the nodes. Positive influence domination has applications in social networks where this parameter is used in \cite{AL18, WWTZ13, DWS14} in order to influence the opinion of the users and individual behaviors in social networks. For example, in a social network with smoking problem, smokers could be exposed to a possible conversion to abstain due to the domination of their friends. 

Simultaneously, Yahiaoui \textit{et al.} have proposed a self-stabilizing algorithm for \textit{minimal global powerful alliance set} called MGPA \cite{YBHK13} which has the same basic concept of MPIDS. A subset $S$ is said \textit{global powerful alliance set} if for each node $v \in V$, the majority of the neighbors of $v$ are in $S$, that is,  $|N[v] \cap S| \geq |N[v] \cap (V-S)|$, where $N[v]=N(v) \cup \{v\}$. Like MPIDS, MGPA can be also considered as total $\frac{1}{2}$-domination. 

Hedetniemi \textit{et al.} have presented a collection of self-stabilizing algorithms to find an \textit{unfriendly partition} into two dominating sets $R$ and $B$ \cite{HHKM13}. A bipartition $\{R,B\}$ is called \textit{unfriendly partition} if for every node $v \in R$, most of neighbors of $v$ are in $B$ and for every node $u \in B$, most of neighbors of $u$ are in $R$. Obviously, every node of $R$ is $\frac{1}{2}$-dominated by $B$ and every node of $B$ is $\frac{1}{2}$-dominated by $R$.
\subsection{Contribution}
Clearly, the above algorithms discuss a domination which is limited by the value of $\alpha = 1/2$, where every node must have at least $50\%$ of the neighbors in the dominating set. However, it will be more comfortable if this value could be changed in order to be more suitable for particular cases where other values are required. For some cases  $50\%$ can represent a bad choice while other values like $20\%$ $(\alpha=0.2)$ or $70\%$ $(\alpha=0.7)$ may be more useful. Using the general parameter of $\alpha$-domination will be more interesting for many practical cases.  

In this paper, we propose a self-stabilizing algorithm to find the \textit{minimal $\alpha$-dominating set} called $\alpha$-MDS. Analysis shows that our algorithm stabilizes in $2n$ moves, namely a complexity of $O(n)$ using the expression distance-2 model under central daemon. Note that expression distance-2 model is a variant of distance-two model. Using the transformer of \cite{Tur12}, $\alpha$-MDS can be converted to $\alpha$-MDS$^{D}$ that converges in $O(nm)$ under distributed daemon and distance-one model. Table 1 summarizes main formal results of $\alpha$-MDS, MPIDS, MGPA and UNFIENDLIER. 

 
\begin{table}
\label{table1}
\centering
\caption{Self-stabilizing algorithms on $\alpha$-domination problem.}
\begin{tabularx}{\textwidth}{X|X|X|X} \hline

Algorithm							& Daemon	  &	Complexity	&$\alpha$ values	\\ \hline\hline
MPIDS \cite{WWTZ13}				& Central   & $O(n^{2})$  &$\alpha = 1/2$\\
\hline
Unfriendlier \cite{HHKM13}	& Central   & $O(n^{3}m)$ &$\alpha = 1/2$	
\\ \hline	
MGPA \cite{YBHK13}			&Distributed & $O(nm)$	    &$\alpha = 1/2$\\ \hline
$\alpha$-MDS				 		&Distributed & $O(nm)$		&$0 < \alpha \leq 1$		\\ 
\hline\end{tabularx}
\end{table}
\section{Model and Terminology}
Generally, networks or distributed systems are represented by simple undirected graphs. Let $G = (V, E)$ be a connected graph where $V$ is the set of nodes and $E$ is the set of edges. For a node $v \in V$,  the open neighborhood of $v$ is defined as $N(v)=\{u \in V : vu \in E\}$, where the degree of $v$ is $d(v) = |N(v)|$. $N[v] = N(v) \cup {v}$ denotes the closed neighborhood of $v$. In this paper, we use $neighborhood$ to indicate the open neighborhood.

\newtheorem{def1}{Definition}
\begin{def1}
A subset $S \subseteq V$ is a dominating set if for every node $v \in V-S$, there exists a node $u \in S$ such that $v$ is adjacent to $u$ \cite{HHS98}.
\end{def1}


The set $N_{S}(v)$ defines the neighbors of $v$ in $S$ \textit{i.e.} $N_{S}(v)=\{u \in S : vu \in E\}$ and $N_{V-S}(v)$ represents neighbors of $v$ in $V-S$. Consequently, $N(v)=N_{S}(v)\cup N_{V-S}(v)$.

An algorithm is self-stabilizing if it will be able to (1) reach a global correct configuration called $legitimate$ and still in the legitimate state ($closure$) (2) during a finite time after it has started from an unknown configuration. 
To show that an algorithm is self-stabilizing, it is sufficient to prove its $closure$ for the legitimate configuration and its $convergence$ to achieve the desired configuration in a finite time. So, a self-stabilizing algorithm guarantees to converge to the legitimate configuration even if there is any possible transient faults. Also, an algorithm is called $silent$ if in a legitimate state, there is no enabled nodes. Obviously, if an algorithm is silent, the closure is trivially satisfied. 

In a \textit{uniform} self-stabilizing system, all the nodes execute the same collection of rules having the form $\textbf{if}$ $guard$ $\textbf{then}$ $statement$ (written as: $guard \longrightarrow statement$). Nodes have also the same local variables that describe their $state$.  The \textit{guard} is a (or a collection of) boolean expression. Once a guard of any node is true, the corresponding statement must be executed (an action on the node's state). 
Thus, the $state$ of every node is updated (modified or not) by the node itself using at least one of its own rules. 
Each node has a partial view of the distributed system ($i.e.$ $guard$ which consists of boolean expressions) on (1) its state and the states of its neighbors (called distance-one model) or (2) its state and the states of its neighbors and the states of the neighbors of its neighbors (called distance-two model). A rule is said $enabled$ if the guard is evaluated to be true. A node will be enabled if at least one of its rules is enabled. Executing the statement of the enabled rule by the node is called a $move$. A move allows updating the state (local variables) of the node in order to be in harmony with its neighborhood.

\subsection{Execution Model}
The execution of self-stabilizing algorithms is managed by a daemon (scheduler) that selects one of the enabled nodes to move from a configuration to another configuration. Two types of daemons are widely used in self-stabilization literature: central and distributed daemons. In the central daemons, one enabled node is selected among all the enabled nodes to be moved. However, in the distributed daemons, a subset of nodes are selected among the set of enabled nodes to make a move simultaneously. A particular case is distinguished for distributed daemons \textit{i.e.} the $synchronous$ daemon where all the enabled nodes are selected to move simultaneously. Indeed, $Unfair$ $distributed$ $daemon$ is the most used scheduler in self-stabilizing literature, wherein any subset of the enabled nodes can make their moves simultaneously. A $fair$ daemon is a scheduler that selects the same enabled node continuously between configurations transition. Otherwise, the daemon is $unfair$ where it can delay the node move if there are other enabled nodes which allows to guarantee the convergence to the global legitimate configuration. A detailed taxonomy of the daemons variants can be found in \cite{DT11}.

\subsection{Transformers}
Generally, it is easy to prove the stabilization of an algorithm working under hypotheses like central daemon and expression distance-2 model. However, algorithms working under distributed daemon and distance-one model are more difficult to prove, although they are more suitable for real applications.

A common approach, known in literature \cite{Tur12,GS13}, allows converting a self-stabilizing algorithm $A$ which operates under a given hypotheses to a new self-stabilizing algorithm $A^{T}$, such that $A^{T}$ operates under other hypotheses. This transformation guarantees that the two algorithms obtain the same legitimate configuration. 

Note that in the case of distributed daemon, it is not allowed for neighbors to execute simultaneously a move at the same time. This is achieved by using unique identifiers for nodes. At the same round, we can allow solely to the node having the higher identifier (among the enabled neighbors) to make a move. Whereas, the remainder neighbors still enabled in the next round.


\section{Minimal $\alpha$-Dominating Set}
In this section, we present a self-stabilizing algorithm for finding minimal $\alpha$-dominating set, we call $\alpha$-MDS. 
First, we give defintion of $\alpha$ dominating set:

\begin{def1}
Let $S$ be a subset of $V$ and $0 < \alpha \leq 1$. $S$ is called \textit{ $\alpha$-dominating set} if for every node $v \in V-S$, $\frac{|N_{S}(v)|}{|N(v)|} \geq \alpha$. $S$ is minimal if no proper subset of $S$ is $\alpha$-dominating set. Every node in $V-S$ is called $\alpha-dominated$.
\end{def1}

In algorithm \ref{alg1}, each node $v$ maintains a local variable $state$ and two expressions $exp1$ and $exp2$. The value of $state$ can be $In$ or $Out$. It is clear that $state$ is used to express if any node belongs to the $\alpha$-dominating set or not. Therefore, $\alpha$-dominating set is defined as $S=\{v \in V : v.state = In \}$. $exp1$ is used to check if every node of $V-S$ is $\alpha$-dominated. $R1$ shows that every node in $V-S$ which is not dominated must convert its state from $Out$ to $In$. Consequently, $R1$ ensures that every node of $V-S$ will be $\alpha$-dominated. $R2$ is used to verify the minimality of $S$. Every node in $S$ that can leave $S$ without affecting the constraint $\alpha$-dominated of its neighbors in $V-S$ and still itself $\alpha$-dominated, will leave $S$ because it preserves the correct configuration in its neighborhood. Observe that $exp2$ is used in this situation: when a node $v$ wants to leave $S$, all the neighbors $w \in  N_{V-S}(v) $ must still respect the $\alpha$-domination condition after the leaving of $v$.

					

\begin{figure*}
\centering{ \fbox{
\begin{minipage}[t]{150mm}
\footnotesize
\resetline
\begin{tabbing}

{\it Each node $v$ checks and executes (with infinite loop) the following } \\
{\bf Expre}\={\bf ssions:}\\
\>-----------------------------------------------------------------------------------------------------------------------  \\
\>\ 	{\bf exp1}::~ $\frac{|N_{S}(v)|}{|N(v)|}$   \\
\>\	{\bf exp2}::~\=  $\frac{|N_{S}(v)|-1}{|N(v)|}=exp1-\frac{1}{|N(v)|}$  \\
\>-----------------------------------------------------------------------------------------------------------------------  \\
{\bf Rules:}\={\bf}\\
\> -----------------------------------------------------------------------------------------------------------------------  \\

\>\  {\bf R1:} $v.state=Out \wedge v.exp1 < \alpha   \longrightarrow v.state=In$\\
\>\  {\bf R2:} $v.state=In \wedge (v.exp1 \geq \alpha) \wedge   (  \forall w \in  N_{V-S}(v) : w.exp2 \geq \alpha)\longrightarrow v.state=Out$\\

\>-----------------------------------------------------------------------------------------------------------------------  \\

\end{tabbing}
\normalsize
\end{minipage}
} \caption{$\alpha$-MDS self-stabilizing algorithm}
\label{prot3}}
\end{figure*}

\subsection{Closure}

\newtheorem{lem}{Lemma}
\begin{lem}
Once all the nodes are not enabled, the set $S$ is a minimal $\alpha$-dominating set.
\label{lem1}
\end{lem}

\begin{proofL}
We prove that : (a) every node $v$ of $V-S$ is $\alpha$-dominated and (b) $S$ is minimal.
     
(a) For every node $v$ out of $S$,  $ v.exp1$ must be $\geq \alpha$ because $R1$ is not enabled. Hence, each node out of $S$ is $\alpha$-dominated. 

(b) Suppose that all the nodes of $S$ are not enabled and there exists a node $v \in S$ such that $S' = S-\{v\}$ is minimal $\alpha$-dominating set. Thus, $\frac{|N_{S'}(v)|}{|N(v)|} \geq \alpha$ (by definition) which implies that $\frac{|N_{S}(v)|}{|N(v)|} \geq \alpha$ because $N_{S'}(v)=N_{S}(v)$ . Since all the nodes of $S$ are not enabled ($R2$ is not enabled for $v$) and $v.exp1 \geq \alpha$, there exists a node $w \in N_{V-S}(v)$ such that $w.exp2 < \alpha$, so $\frac{|N_{S}(w)|-1}{|N(w)|}< \alpha$. After moving $v$ from $S$ to $V-S$, the number of neighbors of $w$ having state $In$ will decrease by one \textit{i.e.}  $|N_{S'}(w)|=|N_{S}(w)|-1$. Thus $\frac{|N_{S}(w)|-1}{|N(w)|}< \alpha$ becomes $\frac{|N_{S'}(w)|}{|N(w)|}< \alpha$. This is a contradiction with the definition that every node out of the dominating set must be $\alpha$-dominated.
\end{proofL} 

\subsection{Convergence and Complexity Analysis}
\label{Convergence and Complexity Analysis }

\begin{lem}
If any node $w \in (V-S)$ has $w.exp1 \geq \alpha$, the value $w.exp1 $ still greater or equal than $\alpha$ and cannot be down less than $\alpha$.
\label{lem2}
\end{lem}

\begin{proofL}
Let $w$ be a node from $V-S$ such that $w.exp1 \geq \alpha$. The value $w.exp1$ can be decrease by one way which is: if any neighbor $v$ of $w$ changes its state from $In$ to $Out$. However, R2 imposes that when $v$ moves from $In$ to $Out$ , all its neighbors in $V-S$ including $w$ must have $exp2 \geq \alpha$ \textit{i.e.} $\frac{|N_{S}(w)|-1}{|N(w)|} \geq \alpha$ \footnote{If at least one of $v$ neighbors: $w' \in S'$ has $w'.exp2 < \alpha$, then R2 cannot be enabled and $v$ remains in $S$, thus $w.exp1$ still has $\geq \alpha$.}. Suppose $S_{2}$ is the new dominating set after $v$ leaves $S$. Thus, $|N_{S_{2}}(w)|=|N_{S}(w)-1|$. So, $\frac{|N_{S}(w)|-1}{|N(w)|} \geq \alpha$ becomes $\frac{|N_{S_{2}}(w)|}{|N(w)|} \geq \alpha$. Hence, $w.exp1$ remains $\geq \alpha$.
\end{proofL}

\begin{lem}
Once a node leaves $S$, it cannot reach $S$ again.
\label{lem3}
\end{lem}
\begin{proofL}
Since any node leaves $S$ with $exp1 \geq \alpha$, R1 cannot be enabled again according Lemma \ref{lem2}.   
\end{proofL}

\begin{lem}
Every node executes at most R1 then R2 which allows algorithm 1 to terminate in the worst case at $2n$ moves under the expression distance-2 model using unfair central daemon.
\label{lem4}
\end{lem}

\begin{proofL}
It follows from Lemma \ref{lem3}.
\end{proofL}

\begin{theorem}
$\alpha$-MDS is a silent self-stabilizing algorithm giving Minimal $\alpha$-Dominating Set in finite time not exceeding $O(n)$ moves under expression distance-2 model using unfair central daemon.
\label{th1}
\end{theorem}

\begin{proofT}
The proof follows from Lemma \ref{lem1} and  Lemma \ref{lem4}. 
\end{proofT}

After proving the stabilization of algorithm $\alpha$-MDS under the central daemon and expression distance-2 model, we use the transformer proposed by \cite{Tur12} that gives another self-stabilizing algorithm $\alpha$-MDS$^{D}$. This later is executable under distributed daemon and distance-one model.

\begin{theorem}
$\alpha$-MDS$^{D}$ gives a minimal $\alpha$-dominating set and stabilizes in $O(nm)$ moves in the distance-one model under unfair distributed daemon .
\label{th2}
\end{theorem} 

\begin{proofT}
Using Theorem \ref{th1}, the proof follows from Theorem 18 of \cite{Tur12} , where $m$ is the number of edges. 

\end{proofT}


\section{Simulation and Experimental results}
\begin{figure}
\centering
\includegraphics[height=2.5in, width=3in]{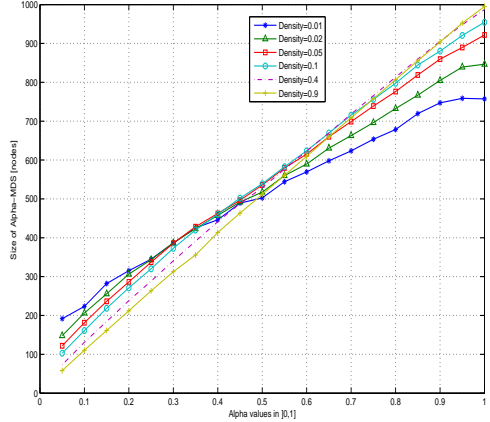}
\caption{Cardinality of $\alpha$-MDS according to $\alpha$ on graphs with 1000 nodes.}
\label{fig1}
\end{figure}In this section, simulation tests are carried out to evaluate $\alpha$-MDS on two levels. First, we attempt to observe the behavior of $\alpha$-MDS according values of $\alpha$ in $]0,1]$. Secondly, we compare the performance of $\alpha$-MDS with other known algorithms, namely MGPA, MTDS and MKDS that have been proposed by \cite{YBHK13}, \cite{BYK14} and \cite{Tur12,WWTZ12}, respectively. For each level, two parameters of efficiency are used: the cardinality of the dominating set and the time of convergence. Clearly, the efficient algorithm is that who gives the smallest dominating set and/or converges more quickly. Recall that we have selected these algorithms because they are the solely proposed in literature under expression distance-2 model.  It is important to mention that this is the first work implementing the expression model. The choice of the daemon and the graphs follows the implementation of Lukasz Kuszner \cite{Kus05} whereby we have used a central daemon and generated arbitrary graphs with different density having orders from 1000 nodes to 10000 nodes. For each size of graphs, we have carried out 5 executions and then we have taken the average value.  

\begin{figure}
\centering
\includegraphics[height=2.5in, width=3.0in]{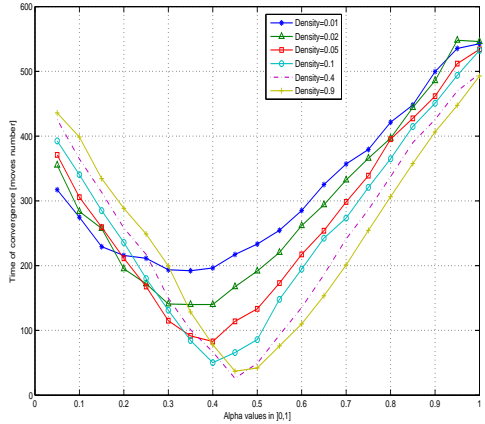}
\caption{Convergence according to $\alpha$ on graphs with 1000 nodes.}
\label{fig2}
\end{figure}

Figures \ref{fig1} and \ref{fig2} show experiments performed on $\alpha$-MDS only. In this case, we try to understand the behavior of $\alpha$-MDS according values of $\alpha$ in $]0,1]$. Figure \ref{fig1} illustrates that whatever the graph density, the $\alpha$-MDS size grows proportionally whith $\alpha$ values. However, density of graph has some impact on the cardinality of the dominating set. For high density ($=0.9$) where graphs are close to be complete, relation is clear: $\alpha \simeq cardinality$ like an equation of a line $x=y$, where $0 < \alpha \leq 1$. Once density begins to be down, curves of $\alpha$-MDS size starts to deviate from the line $x=y$ especially on the extremities of the interval $]0,1]$. The worst deviation from $x=y$ is represented by the curve of the smallest density $=0.01$ where for $\alpha=0.05$ the cardinality of $\alpha$-MDS is $20\%$ and for  $\alpha=0.95$ the cardinality is $75\%$. Theses results are important from a point of view application. Reducing the value of $\alpha$ as possible gives a lower cardinality of the dominating set which is very practical in the reality. For example, for a given problem, if we want a small dominating set of nodes, it will be sufficient to set $\alpha$ as small as possible.

\begin{figure}
\centering
\includegraphics[height=2.5in, width=3.0in]{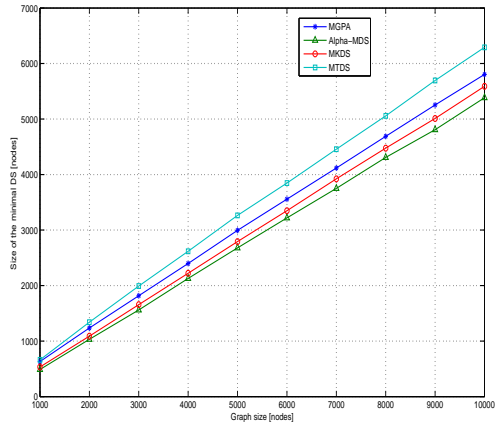}
\caption{Cardinality of DS for each algorithm.}
\label{fig3}
\end{figure}

Figure \ref{fig2} shows the necessary time to converge to the stable configuration according values of $\alpha$. 
Theoretically, we have proved in section \ref{Convergence and Complexity Analysis } that the number of moves cannot exceed $2n$ moves which is confirmed by the experiments where the number of moves is always less then $6n/10$. However, it is clear through Figure \ref{fig2} that $\alpha$-MDS needs more time (number of moves) on the extremities of $]0,1]$ while it converges quickly in the middle of this area.

\begin{figure}
\centering
\includegraphics[height=2.5in, width=3.0in]{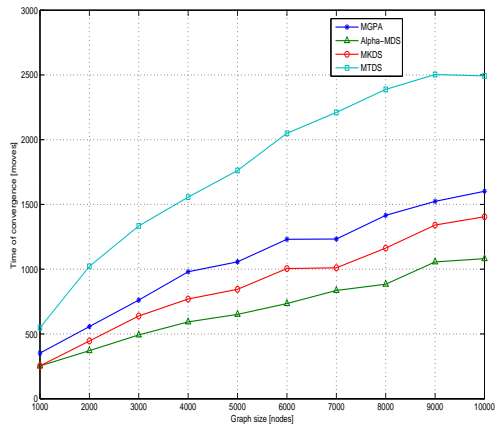}
\caption{Convergence comparison.}
\end{figure}
\label{fig4}

Figures \ref{fig3} and \ref{fig4} illustrates comparison experiments between our algorithm $\alpha$-MDS, MGPA \cite{YBHK13}, MTDS \cite{BYK14} and MKDS \cite{Tur12,WWTZ12}. 
Note that we have take $\alpha=0.5$ in this case. Recall that we have shown in section \ref{Related works} that MGPA is a particular case of $\alpha$-MDS where $\alpha=0.5$. However, we have used the generalized version of MTDS and MKDS where $k$ is calculated according the density $d$ as follow: $k=\frac{n*d}{2}$ which is approximately the half degree for each node. 
Results show that $\alpha$-MDS outperforms other algorithms in both sides $i.e.$ giving smallest cardinality of the dominating set and stabilizing more quickly. Probably, other factors could affect the simulation results. For example, our algorithm impose its constraint of $\alpha$-domination just on the nodes out of the dominating set. However, in MTDS and MGPA the imposed constraint must be respected for all the nodes which needs more time to stabilize and gives higher cardinality.

\section{Conclusion}
A self-stabilizing algorithm is proposed in this paper for the parametric $\alpha$-domination called minimal $\alpha$-dominating set. The algorithm is studied in both sides theoretical and experimental, where it is proved that our algorithm converges in $O(nm)$ using a distributed daemon. Although the complexity is the same regarding other algorithms, experimental simulations shows that $\alpha$-MDS is more efficient.
The algorithm is useful for real-life use particularly for security and health applications.

\bibliographystyle{abbrvurl}
\bibliography{mybibfile}


\end{document}